\documentclass[aps,prl,twocolumn,groupedaddress,showpacs]{revtex4}

\usepackage{graphicx}

\begin{document}

\title{Realizing the strongly correlated $d$-Mott state in a fermionic cold atom optical lattice}
\author{Michael R. Peterson, Chuanwei Zhang, Sumanta Tewari, S. Das Sarma}
\affiliation{Condensed Matter Theory Center, Department of Physics, University of
Maryland, College Park, MD 20742}
\date{\today}

\begin{abstract}
We show that a new state of matter,
the $d$-Mott state (introduced recently by H. Yao, W. F. Tsai, and
S. A. Kivelson, Phys.  Rev. B 76, 161104 (2007)), which is characterized
by a non-zero expectation value of a local plaquette operator embedded in an
insulating state, can be engineered using ultra-cold atomic fermions in
two-dimensional double-well optical lattices. We characterize and analyze
the parameter regime where the $d$-Mott state is stable. We predict the
testable signatures of the state in the time-of-flight measurements.
\end{abstract}

\maketitle

Experimental realizations of degenerate Bose~\cite{Bloch1,Zoller1} and
Fermi gases~\cite{Modugno,Kohl,Chin,Esslinger} in cold atom optical
lattices has led to increasing interaction between the atomic and
condensed matter physics communities. In fact, these atomic systems
have become a proverbial \textquotedblleft
playground\textquotedblright\ where, in principle, custom
Hamiltonians~\cite{Bloch2,Lewenstein,Duan} can be made to order,
enabling optical lattice emulations of strongly correlated condensed
matter phenomena. Particularly exciting, in this context, is the
creation of a fermionic Hubbard model in ultra-cold atomic
systems, since the Hubbard model is a paradigm for theoretical studies
of strong correlations. The two-dimensional fermionic Hubbard model, a
model woefully difficult to \textquotedblleft solve\textquotedblright
, is of great interest since it is thought to hold the key to
understanding high-temperature (high-$T_{C}$)
superconductivity~\cite{Lee}. In fact, the cold atom systems allow for
versatile tuning of the Hamiltonian parameters where the hopping (or
tunneling) matrix element between nearest neighbor sites (through
laser intensity tuning), the on-site interaction between particles
(from attractive to repulsive by tuning the Feshbach resonance), and
the dimensionality (from one to three) of the model can all be varied
and controlled~\cite{Bloch2}.  Such versatility in tuning the
Hamiltonian makes cold atom systems particularly attractive in
studying novel quantum phases and transitions between them.

The possible $d$-wave character of the ground state in the Hubbard
model is an important concept in relation to the high-$T_{C}$ cuprate
superconductors~\cite{Lee}. Studying the Hubbard model experimentally
in the cold atomic gases could, in principle, be an effective and
accurate analog simulation of high-$T_C$ superconductivity.
Eventually it would be very important to test whether or not the
fermionic Hubbard model, in its full generality, allows for $d$-wave
superfluidity~\cite{Hofstetter,Trebst,Bloch3}, a controversial
conjecture not yet settled theoretically.  In this work, we study the
model in a controlled limit in which there is (i) an exactly solvable
point where a true new state of matter, the so-called $d$-Mott state
recently discovered by Yao, et al.~\cite{Yao} exists as the ground state
of the checkerboard Hubbard model (see below), and (ii) a parameter
which can be varied to destroy the $d$-Mott state. The $d$-Mott state~\cite{Yao}
is a special Mott insulating state with a non-zero expectation value
($-1$) of a local plaquette operator, $D_P$ (see below), which
cyclically rotates the sites by an angle $\pi /2$.  In other words,
the state is an insulator with a local plaquette $d$-wave symmetry--in
analogy with the $d$-wave superconductor--characterized by a non-zero
expectation value $\langle D_P\rangle$.  Realizing this state in an
optical lattice has its own intrinsic appeal because it is a true new
state of matter not adiabatically connected to any other known
insulating states~\cite{Yao}. Furthermore, since this is the ground
state arising out of an exact solution of the Hubbard model in a
controlled limit, it would serve as an important benchmark for the
accuracy of experiments aimed at studying the general Hubbard
model. For this purpose, we propose and analyze how the distinctive
signatures of the $d$-Mott state could be observed in the usual
time-of-flight measurements.  An alternative proposal was provided in Ref.~\cite{Yao}.

\begin{figure}[t]
\begin{center}
\includegraphics[scale=0.35]{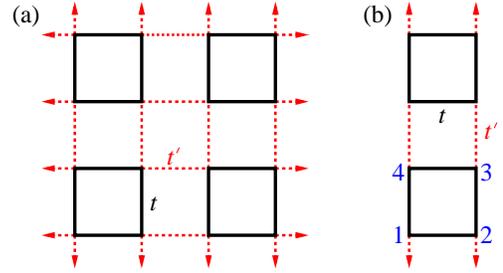}
\end{center}
\caption{(Color online) (a) Two-dimensional checkerboard Hubbard model.
Solid (dashed) bonds represent hopping amplitudes $t$ ($t^{\prime }$). (b)
\textquotedblleft Ladder\textquotedblright\ cluster we exactly diagonalize
with periodic boundary conditions imposed in the \textquotedblleft $y$
\textquotedblright -direction and open boundary conditions in the
\textquotedblleft $x$\textquotedblright -direction. The numbers label the
sites of a single plaquette.}
\label{lattice}
\end{figure}

We consider a system of fermionic atoms ($^{40}K$) where the two
atomic internal hyperfine states, $\left\vert
F=9/2,m_{F}=-9/2\right\rangle $ and $\left\vert
F=9/2,m_{F}=-7/2\right\rangle $, are taken as the effective spin
states $\sigma =\uparrow ,\downarrow $. The atoms are loaded into a
two dimensional superlattice, produced by superimposing a long and a
short period lattice in the $x$- and $y$-direction such that an array
of square plaquettes is created~\cite{Bloch4,Porto}. The dynamics of
the atoms can be described by the so-called checkerboard Hubbard
model~\cite{Tsai} defined as
\begin{equation}
H=-\sum_{{\mathbf{r}}{\mathbf{\delta }}\sigma }t_{{\mathbf{r}}}({\mathbf{
\delta }})c_{{\mathbf{r}}+{\mathbf{\delta }}\sigma }^{\dagger }c_{{\mathbf{r}
}\sigma }
+U\sum_{\mathbf{r}}n_{\mathbf{r}\uparrow }n_{\mathbf{r
}\downarrow }\;.  \label{Ham}
\end{equation}
Here, $c_{\mathbf{r}\sigma }\;(c_{\mathbf{r}\sigma }^{\dagger })$ is
the particle destruction (creation) operator at lattice site
$\mathbf{r}$ with spin (or magnetic sublevel) $\sigma $,
$t_{\mathbf{r}}(\mathbf{\delta })$ is the hopping amplitude for a
particle at site $\mathbf{r}$ hopping to site
$\mathbf{r}+\mathbf{\delta }$, where site $\mathbf{r}+\mathbf{\delta
}$ is a nearest neighbor site, $U$ is the usual on-site interaction
energy, and
$n_{\mathbf{r}\sigma}=c^\dagger_{\mathbf{r}\sigma}c_{\mathbf{r}\sigma}$
is the number operator. The
checkerboard Hubbard model is defined by choosing
$t_{\mathbf{r}}(\mathbf{\delta })=t$ when $\mathbf{\delta }$ connects
sites within a square plaquette, and $t_{\mathbf{r}}(\mathbf{\delta
})=t^{\prime }$ when $\mathbf{\delta }$ connects sites between two
neighboring square plaquettes, see Fig.~\ref{lattice}.

When $t^{\prime }=0$ the ground state of this model can be obtained
exactly for all $U/t$ and all band-fillings or densities (dopings) by
solving the single plaquette problem exactly (either by brute
force~\cite{Schumann}, Bethe ansatz~\cite{Lieb} -- four site plaquette
is a one-dimensional ring with four sites -- or through numerical
exact diagonalization) and constructing the full many-plaquette state
as a product state of the plaquette basis states. At exactly half
filling (or zero doping, the case considered throughout this work),
the ground state for $U/t>0$ is an insulator with $d$-wave symmetry on
the plaquette, the so-called $d$-Mott state~\cite{Yao}: the plaquette
state is odd under $\pi /2$ rotations about the plaquette center. 

To measure the symmetry characteristic of the $d$-Mott state on a
plaquette we propose an operator, $D_{P}$, which, when operating on a
state, rotates a plaquette by $\pi /2$ about its center or, alternatively, 
translates each particle by one site along the four-site
\textquotedblleft ring\textquotedblright.  More concretely, for a
particular basis state for demonstrative purposes, the single
four-particle four-site plaquette (such as the lattice labeled in
Fig.~\ref{lattice}(b)) occupation could be
$|n_{1\uparrow}n_{2\uparrow}n_{3\uparrow}n_{4\uparrow};
n_{1\downarrow}n_{2\downarrow}n_{3\downarrow}n_{4\downarrow}\rangle=|1100;0110\rangle$
where sites 1 and 2 are occupied with spin-up particles
and 2 and 3 are occupied with spin-down.  The result of operating
$D_P$ on this state is $D_P|1100;0110\rangle=|0110;0011\rangle$.  The
particle at site 1 moves to 2, 2$\rightarrow $3, 3$ \rightarrow $4,
and 4$\rightarrow $1 (the last through periodic boundary conditions on
the plaquette). It is easy to see that the possible eigenvalues of
$D_P$ are $\pm 1$, $\pm i$ since $(D_P)^{4} =\mathbf{1}$.

\begin{figure}[t]
\begin{center}
\includegraphics[width=6.cm,angle=0]{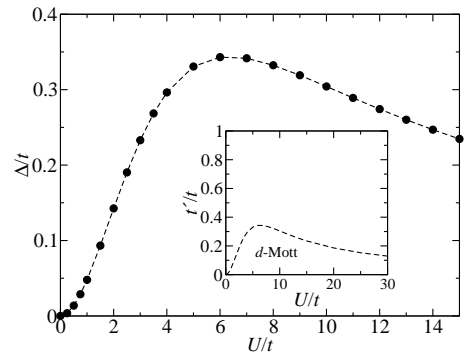}
\end{center}
\caption{The gap $\Delta $ as a function of $U/t$ for a single plaquette ($
t^{\prime }=0$).  The inset shows, approximately, the 
region in $(U/t)$-$(t^\prime/t)$ space where the $d$-Mott state is expected 
to exist.}
\label{gapfig}
\end{figure}

For $t^{\prime }=0$ and $U/t=0$ the ground state is six-fold
degenerate and the operation of $D_P$ yields $\pm 1$ or $\pm i$ for
each degenerate ground state. When $t^{\prime }=0$ and $U/t>0$, the
degeneracy is broken and a gap $\Delta$ opens up.  The
operation of $D_P$ onto the ground state now yields an eigenvalue
equal to $-1$ indicating the existence of $d$-wave symmetry--the state
has a non-zero value of the $d$-Mott operator $\langle
D_P\rangle$.  In Fig.~\ref{gapfig}, we plot the gap $\Delta $ with
respect to $U/t$.  The gap is zero for $U=0$ due to the degeneracy of
the ground state while with finite $U$ the degeneracy is lifted and
the gap becomes nonzero. However, the gap also approaches zero in the
limit $U/t\rightarrow\infty$ where there is no phase coherence between
the atoms at different sites and the ground state is again
degenerate. At a finite $U/t\approx 6$, the gap is maximum.

We now consider what happens when $t^{\prime }\neq 0$. An important
point to note is that $D_P$ does not commute with the Hamiltonian for
$t^{\prime }\neq 0$, therefore eigenstates of $H$ will not be
eigenstates of $D_P$, altough $H$ retains a global $C_4$ symmetry about 
the center of any one plaquette.  However, there are clearly defined regimes for
non-zero values of $t^{\prime} $. As mentioned, for $t^{\prime }=0$
and $U/t>0$, the ground state is gapped and it is an eigenstate of $H$
and $D_P$ (with eigenvalue -1).  For $t^{\prime }\ll \Delta $ the
ground state is approximately a product state of single plaquette
ground states and, hence, remains approximately a simultaneous
eigenstate of $H$ and $D_P$ (by this we mean that the ground state
will have the largest amplitude from basis states that are $D_P$
eigenstates with eigenvalue $-1$). Thus, it is conceivable that the
average value of $D_P$ for any given plaquette will continue to be
nonzero and close to $-1$ for a non-zero range of $ t^{\prime}$. On the other hand, when
$t^{\prime } \sim t$, the Hamiltonian is approximately the standard
Hubbard model.  Thus, the ground state is an essentially equal
superposition of $D_P$ eigenstates, so one expects the value of $\langle D_P\rangle$, 
for any given plaquette, to be zero, since $1+(-1)+i+(-i)=0$.

\begin{figure}[t]
\begin{center}
\includegraphics[width=7.0cm,angle=0]{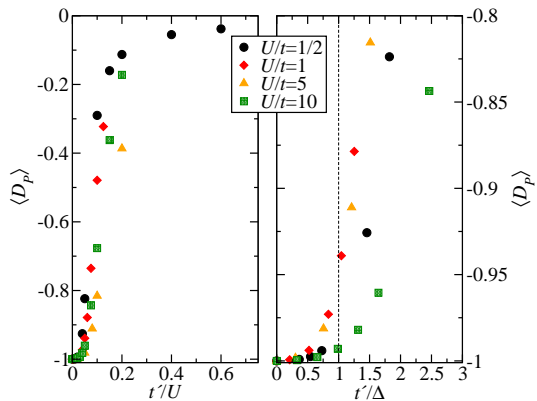}
\end{center}
\caption{(Color online) Ground state expectation value of the $d$-Mott
operator, $\langle D_P\rangle $, as a function of $t^{\prime }/U$ (left)
and $t^{\prime }/\Delta $ (right) for the eight-site square lattice ladder
system for the checkerboard Hubbard model at half filling for 
various values of $U/t$.}
\label{d-mott}
\end{figure}

The question now becomes exactly what happens for $U/t>0$ and
$t^{\prime }\approx \Delta $ when the particles are first able to hop
to the neighboring plaquettes. To answer this question we exactly
diagonalize $H$ for eight particles on an eight-site square lattice
ladder with periodic boundary conditions in the ``$y$"-direction
depicted in Fig.~\ref{lattice}(b) and vary both $U/t$ and $t^{\prime
}/t$ from zero to $t^{\prime }/t=1$ and beyond. (Note that a fully
periodic system--torus--requires 16 sites with a quite large Hilbert
space--$\sim$ 100 million states--which is not necessary for
establishing the qualitative and semi-quantitative conclusions
presented here.)  The $d$-Mott character is examined by calculating
the average $\langle D_P\rangle$. In Fig.~\ref{d-mott} we plot
$\langle D_P\rangle$ as a function of $t^{\prime }/U$ (left panel) and
$ t^{\prime }/\Delta $ (right panel) for various values of $U/t$
(weakly to strongly interacting). From the left panel we see that the
behavior is largely universal when plotted versus $t^{\prime}/U$ and
the value drops well below $-1$ for $t^{\prime }\approx t $. However,
the physics is better elucidated in the right panel. Here it is
clearly seen that once $t^{\prime }$ becomes equal to, and exceeds,
the single plaquette gap $\Delta $ the $d$-Mott order is quickly
lost. Thus, the optimal parameter regime for observing the $d$-Mott
state is at $U/t\approx 6$, where the gap $\Delta $ reaches the
maximum and the $d$-Mott state, with a non-zero expectation value of
the local plaquette operator $D_P$, is most stable against the
inter-plaquette coupling $t^{\prime }$.  See the inset of
Fig.~\ref{gapfig} for a plot showing the region in
$(U/t)$-$(t^\prime/t)$ space where the $d$-Mott state is expected.

We now turn to how, in a cold atomic gas, the $d$-Mott state could be
observed via time-of-flight measurements. First, we suggest a method
to generate the $d$-Mott state in experiments. To begin, the potential
depth of the short period optical lattice is ramped up to a large
value $V_{s}=20E_{R}$ to form a Mott state with one atom per lattice
site~\cite{Esslinger}, where $E_{R}$ is the photon recoil energy. The
potential depth of the long period optical lattice is then ramped up
to $V_{l}=17E_{R}$. In the Mott state, we have $t=t^{\prime }\approx
0$. To create the $d$-Mott state, the potential depth of the short
period optical lattice is then ramped down to a small value (for
instance, $V_{s}=3E_{R}$) to enhance the tunneling $t$ inside the
plaquettes. The potential barrier between plaquettes is
$V_{l}+V_{s}=20E_{R}$, therefore $t^{\prime }\approx 0$. There are now
four atoms in each plaquette, which corresponds to the half filled
case discussed above. Non-zero $t^{\prime }$ can be obtained by
lowering $V_{l}$ and the interaction strength $U$ can be adjusted in
experiments using Feshbach resonance~\cite{Esslinger}. With this
method, various parameter regimes can be reached with different $U/t$
and $t^{\prime }/t$.  For the regime of $U/t\approx 6$, $t^{\prime
}/\Delta \ll 1$ (i.e. $t^{\prime }/t\ll 0.3$ from Fig.~\ref{gapfig}),
a stable $d$-Mott state is obtained. We now address the question of
the experimental observation of this state.

\begin{figure}[t]
\begin{center}
\includegraphics[scale=0.425]{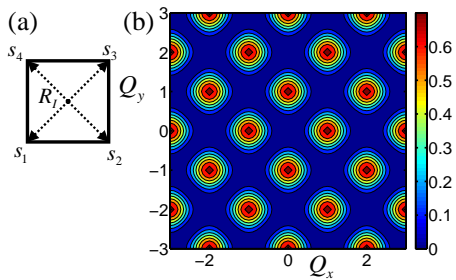}
\end{center}
\caption{(Color online) (a) A single plaquette $I$. (b) Density distribution
of $d$-Mott state in a time of flight experiment. The unit for $Q_{x}$, and $
Q_{y}$ is $\protect\pi /a$. $U/t=5$.}
\label{TOF}
\end{figure}

Interestingly, the phase coherence in each plaquette in the $d$-Mott
state can be experimentally identified in the time-of-flight (TOF)
measurements.  The $d$-Mott state in a single plaquette can be written
as $\left\vert \chi \right\rangle =\sum_{\alpha }f_{\alpha }\left\vert
\alpha \right\rangle$ where $\left\vert \alpha \right\rangle $ is a
basis state such that the action of $D_P$ on a state rotates it to
another basis state and the $d$-wave symmetry of the ground state
wavefunction has been incorporated into the eigen-coefficients
$f_{\alpha }$. In the TOF experiment, we can measure the density of
the spin up and spin down atoms separately. The TOF density
distribution for spin up atoms is
\begin{equation}
\left\langle n_{\uparrow }\left( \mathbf{Q}\left( \mathbf{r}\right) \right)
\right\rangle \propto \sum_{I,I^{\prime }}\psi _{I}^{\ast }\psi _{I^{\prime
}}\left\langle b_{I}^{\dag }b_{I^{\prime }}\right\rangle   \label{density}
\end{equation}
where $\mathbf{Q}\left( \mathbf{r}\right) =m\mathbf{r}/\hbar t$ ($m$
and $t$ being the atomic mass and time of measurement, respectively),
$I$ is the index of plaquettes,
\begin{equation}
\psi _{I}\propto e^{-i\mathbf{Q}\left( \mathbf{r}\right) \cdot \mathbf{R}
_{I}}\Phi \left( \mathbf{Q}\right) \sum_{\alpha }e^{-i\mathbf{Q}\left( 
\mathbf{r}\right) \cdot \sum_{\beta _{\alpha j}=\uparrow }\mathbf{s}_{j}}f_{\alpha }
\label{wave}
\end{equation}
is the wavefunction of spin up atoms in plaquette $I$ after TOF,
$b_{I}^{\dag }$ is the creation operator of the $d$-Mott state in the
plaquette $I$, $ \mathbf{R}_{I}$ is the position of the center of
plaquette $I$, $j=1,2,3,4$ correspond to four lattice sites at each
plaquette (see Fig.~\ref{TOF}(a)) with $\mathbf{s}_{1}=\left(
-a/2,-a/2\right) $, $\mathbf{s}_{2}=\left( a/2,-a/2\right) $, $
\mathbf{s}_{3}=\left( a/2,a/2\right) $, $\mathbf{s}_{4}=\left(
-a/2,a/2\right) $, $a$ is the lattice spacing, $\beta _{\alpha j}$ is
the spin state of atoms at site $j$ and basis state $\alpha $, and
$\Phi \left( Q\right) $ is the Fourier transform of the Wannier
function at each site. In the $d$-Mott state, we have $ \left\langle
b_{I}^{\dag }b_{I^{\prime }}\right\rangle \propto \delta _{II^{\prime
}}$, yielding
\begin{equation}
\left\langle n_{\uparrow }\left( \mathbf{Q}\left( \mathbf{r}\right) \right)
\right\rangle \propto \left\vert \sum_{\alpha }e^{-i\mathbf{Q}\left( \mathbf{
r}\right) \cdot \sum_{\beta _{\alpha j}=\uparrow }\mathbf{s}_{j}}f_{\alpha
}\right\vert ^{2}.  \label{den}
\end{equation}

In a single plaquette, we find that the $d$-Mott ground state can be
written as
\begin{eqnarray}
|\chi_P\rangle &=& \sum_{i=1}^4(-D_P)^{i-1}[(\lambda-\gamma)|1100;0011\rangle \nonumber\\
&&+\gamma|1100;1100\rangle
- \xi\{|1100;1010\rangle+|1100;0101\rangle\}]\nonumber\\
&&+\sum_{i=1}^2(-D_P)^{i-1}[\xi\{-|1010;1100\rangle-|1010;0110\rangle\nonumber\\
&&+|1010;1001\rangle+|1010;1010\rangle\}\nonumber\\&&+2(\lambda-\gamma)|1010;0101\rangle]
\label{wave2}
\end{eqnarray}
where $\lambda $, $\gamma $ and $\xi $ are parameters determined by
$(U,t,t^\prime)$ and satisfy normalization ($16\xi ^{2}+12\left( \lambda -\gamma
\right) ^{2}+4\gamma ^{2}=1$).  When $U\rightarrow \infty $, $\lambda
\rightarrow \sqrt{3}/6$, $\xi \rightarrow 0$, $\gamma \rightarrow 0$
and the state becomes
\begin{eqnarray}
|\chi_P\rangle &\rightarrow &\frac{\sqrt{3}}{6}(
|1100;0011\rangle - |0110;1001\rangle + |0011;1100\rangle\nonumber\\&&-|1001;0110\rangle)
+\frac{\sqrt{3}}{3}(|1010;0101\rangle\nonumber\\&&-|0101;1010\rangle)
\label{wave3}
\end{eqnarray}
which has no double occupancy.  Under the operation of $D_P$ onto
$|\chi_P\rangle$ (either Eq.~\ref{wave2} or Eq.~\ref{wave3}), it is
seen that an eigenvalue of $-1$ is obtained.  Using the wavefunction
(Eq.~\ref{wave2}) and Eq. (\ref{den}), we find
\begin{equation}
\left\langle n_{\uparrow }\left( \mathbf{Q}\left( \mathbf{r}\right)
\right) \right\rangle \propto 4\lambda ^{2}\left\vert \cos \left(
Q_{x}a\right) -\cos \left( Q_{y}a\right) \right\vert ^{2}.
\label{den2}
\end{equation}
Note that the ground state of $H$ (Eq.~\ref{Ham} with finite $t^\prime$) is
approximately the $d$-Mott state when $\langle D_P\rangle\approx -1$
and, hence, it is approximately described by a product state of single
plaquette ground states (Eq.~\ref{wave2}) with a TOF density 
distribution given by Eq.~\ref{den2}.

In Fig.~\ref{TOF}(b), we plot the density distribution for the
$d$-Mott state in TOF measurements with the image showing interference
peaks at $ Q_{x}+Q_{y}=N\pi /a$, where $N$ is an odd integer and $Q_x$ and 
$Q_y$ are multiples of $\pi/a$. We want
to emphasize that these peaks disappear in the TOF for a pure Mott
state where one should observe a flat background, and this would be a
way to contrast the present state from a regular Mott insulator.  Further, the
behavior of $\langle D_P\rangle$ as a function of $t^\prime$
(Fig.~\ref{d-mott}) is exactly indicative of the ``$d$-Mott''-ness of
the ground state of $H$ with a TOF signature shown in
Fig.~\ref{TOF}(b).  Therefore, monitoring the TOF interference peaks
while varying $U/t$ and $t^\prime/t$ would be an effective way to
observe the $d$-Mott state.

\begin{figure}[t]
\begin{center}
\includegraphics[width=5.0cm,angle=0]{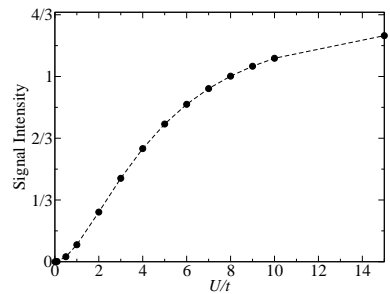}
\end{center}
\caption{Signal intensity $4\protect\lambda ^{2}$ versus $U/t$. }
\label{TOF2}
\end{figure}

In Fig.~\ref{TOF2}, we plot the coefficient $4\lambda ^{2}$ with
respect to the onsite energy $U$ and find that the signal strength
increases reaching 4/3 for $U\rightarrow \infty $. However, the
$d$-Mott state signal may not be observable for large $U$ because the
gap $\Delta $ is small at large $U$ (see Fig.~\ref{gapfig}), where a
small inter-plaquette tunneling $t^{\prime }$ may mix the $d$-Mott
state with other states (see Fig.~\ref{d-mott}(b)), destroying the
order and removing the interference signal. This is expected because
for $U\rightarrow \infty $, the system is in a pure Mott state where
no interference peaks are expected.  Another possible roadblock in the
path of observing the interference pattern is that the temperature of
the system must also be below $\Delta$ so that thermal fluctuations do
not destroy the $d$-Mott order.  But using $U/t\approx$5-6 and keeping
$k_BT\ll\Delta$ should lead to the experimental observation of the
$d$-Mott state.

To conclude, we have described a scenario where a new state of
matter, the $d$-Mott insulator, could be
observed in a fermionic optical lattice via TOF measurements.
Furthermore, we have provided semi-quantitative parameter values where
the state should most likely be seen ($U/t\approx6$ and
$t^\prime/t<0.3$).  This could also serve as a rigorous benchmark in
further experimental studies toward the realization of the general
two-dimensional square lattice Hubbard model in relation to its
utility in the study of high-$T_C$ cuprate superconductors.  We
believe that the direct TOF observation of the hitherto-unobserved
$d$-Mott quantum insulator in a fermionic optical lattice would go a
long way in establishing cold atomic gases as a useful optical lattice
emulator of novel strongly correlated phases.

This work is supported by ARO-DARPA.

\end{document}